\documentclass{appolb}
\usepackage{wrapfig,float,slashed}
\usepackage{amsmath,dsfont}
\usepackage{amssymb,epsfig,graphicx}
\usepackage[table]{xcolor}
\usepackage{epstopdf}
\usepackage[normalem]{ulem}
\usepackage{ragged2e}
\usepackage{lipsum}
\usepackage{siunitx}
\usepackage[font=small,skip=1pt]{caption}
\allowdisplaybreaks
\usepackage{pifont}

\usepackage[colorlinks=true]{hyperref}
\usepackage{tikz}
\usetikzlibrary{arrows,positioning,shapes,mindmap}
\usetikzlibrary{decorations.pathmorphing}
\usetikzlibrary{decorations.markings}
\newcommand{\nc}{\newcommand}
\nc{\non}{\nonumber}
\nc{\hc}{\hbox {H.c.}}
\nc{\noi}{\noindent}
\nc{\barx}{\bar{x}}
\nc{\pbarn}{\;\hbox {pb}}
\nc{\fbarn}{\;\hbox {fb}}
\def\comsp{\@ifnextchar ,\relax{\@ifnextchar\ \relax{\ \relax}}}

\nc{\hsp}{\hspace{0.5cm}}
\nc{\lsp}{\hspace{1cm}}
\nc{\Lsp}{\hspace{2cm}}
\nc{\LLsp}{\lsp\lsp}
\nc{\lra}{\longrightarrow}
\nc{\p}{\prime}
\nc{\sgn}{\text{sgn}}
\nc{\arccot}{\text{arccot}}
\nc{\ph}{\varphi}
\nc{\co}{{\cal O}}
\nc{\beq}{\begin{equation}}  \nc{\eeq}{\end{equation}}
\nc{\bea}{\begin{eqnarray}}  \nc{\eea}{\end{eqnarray}}
\nc{\baa}{\begin{array}}     \nc{\eaa}{\end{array}}
\nc{\bit}{\begin{itemize}}   \nc{\eit}{\end{itemize}}
\nc{\ben}{\begin{enumerate}} \nc{\een}{\end{enumerate}}
\nc{\bce}{\begin{center}}    \nc{\ece}{\end{center}}
\nc{\bpm}{\begin{pmatrix}}   \nc{\epm}{\end{pmatrix}}
\nc{\bvt}{\begin{verbatim}}  \nc{\evt}{\end{verbatim}}
\def\lsim{\mathrel{\raise.3ex\hbox{$<$\kern-.75em\lower1ex\hbox{$\sim$}}}}
\def\gsim{\mathrel{\raise.3ex\hbox{$>$\kern-.75em\lower1ex\hbox{$\sim$}}}}
\def\udots{\mathinner{\mkern1mu\raise1pt\vbox{\kern7pt\hbox{.}}\mkern2mu\raise4pt\hbox{.}\mkern2mu\raise7pt\hbox{.}\mkern1mu}}

\def\gev{\;\hbox{GeV}}
\def\tev{\;\hbox{TeV}}
\def\eq{\hbox{Eq.}~}
\def\mst{M_\ast}
\def\mpl{M_{\text{Pl}}}
\definecolor{agray}{rgb}{0.95, 0.95, 0.99}
\newcommand\fverb{\setbox\fverbbox=\hbox\bgroup\verb}
\newcommand\fverbdo{\egroup\medskip\noindent%
			\fbox{\unhbox\fverbbox}\ }
\newcommand\fverbit{\egroup\item[\fbox{\unhbox\fverbbox}]}
\newbox\fverbbox

\makeatletter
\renewcommand{\boxed}[1]{\textcolor{black}{%
\tikz[baseline={([yshift=-0ex]current bounding box.center)}] \node [rectangle, minimum width=0ex,draw] {\normalcolor\m@th$\displaystyle#1$};}}
 \makeatother

\begin{document}
\pgfdeclaredecoration{complete sines}{initial}
{
    \state{initial}[
        width=+0pt,
        next state=sine,
        persistent precomputation={\pgfmathsetmacro\matchinglength{
            \pgfdecoratedinputsegmentlength / int(\pgfdecoratedinputsegmentlength/\pgfdecorationsegmentlength)}
            \setlength{\pgfdecorationsegmentlength}{\matchinglength pt}
        }] {}
    \state{sine}[width=\pgfdecorationsegmentlength]{
        \pgfpathsine{\pgfpoint{0.25\pgfdecorationsegmentlength}{0.5\pgfdecorationsegmentamplitude}}
        \pgfpathcosine{\pgfpoint{0.25\pgfdecorationsegmentlength}{-0.5\pgfdecorationsegmentamplitude}}
        \pgfpathsine{\pgfpoint{0.25\pgfdecorationsegmentlength}{-0.5\pgfdecorationsegmentamplitude}}
        \pgfpathcosine{\pgfpoint{0.25\pgfdecorationsegmentlength}{0.5\pgfdecorationsegmentamplitude}}
}
    \state{final}{}
}
\tikzset{
fermion/.style={thick,draw=black, line cap=round, postaction={decorate},
    decoration={markings,mark=at position 0.6 with {\arrow[black]{latex}}}},
photon/.style={thick, line cap=round,decorate, draw=black,
    decoration={complete sines,amplitude=4pt, segment length=6pt}},
boson/.style={thick, line cap=round,decorate, draw=black,
    decoration={complete sines,amplitude=4pt,segment length=8pt}},
gluon/.style={thick,line cap=round, decorate, draw=black,
    decoration={coil,aspect=1,amplitude=3pt, segment length=8pt}},
scalar/.style={dashed, thick,line cap=round, decorate, draw=black},
ghost/.style={dotted, thick,line cap=round, decorate, draw=black},
->-/.style={decoration={
  markings,
  mark=at position 0.6 with {\arrow{>}}},postaction={decorate}}
 }
\makeatletter
\tikzset{
    position/.style args={#1 degrees from #2}{
        at=(#2.#1), anchor=#1+180, shift=(#1:\tikz@node@distance)
    }
}
\makeatother
\title{\sc Radius stabilization and dark matter\\
with a bulk Higgs in warped extra dimension%
\thanks{Presented at the XXXIX International Conference of Theoretical Physics ``Matter To The Deepest'', Ustro\'n, Poland, September 13--18, 2015.}%
}
\author{A.~Ahmed$^{1,2}$\footnote{Speaker: \href{mailto:aqeel.ahmed@fuw.edu.pl}{aqeel.ahmed@fuw.edu.pl}}, B.~Grzadkowski$^{1}$, J.~F.~Gunion$^{3}$ and Y.~Jiang$^{4}$
\address{$^{1}$Faculty of Physics, University of Warsaw, Pasteura 5, 02-093 Warsaw, Poland\\
$^{2}$National Centre for Physics, Quaid-i-Azam University Campus,\\
Shahdra Valley Road, Islamabad 45320, Pakistan\\
$^{3}$Department of Physics, University of California, Davis, CA 95616, U.S.A.\\
$^{4}$Niels Bohr International Academy, University of Copenhagen,\\
Blegdamsvej 17, DK-2100
Copenhagen, Denmark}
}
\maketitle
\begin{abstract}
We employ an $SU(2)$ bulk Higgs doublet as the stabilization field in Randall-Sundrum model with appropriate bulk and brane-localized potentials. The gauge hierarchy problem can be solved for an exponentially IR-localized Higgs background field with mild values of fundamental parameters of the 5D theory. We consider an IR-UV-IR background geometry with the 5D SM fields in the bulk such that all the fields have even and odd tower of KK-modes. The zero-mode 4D effective theory contains all the SM fields plus a stable scalar, which serves as a dark matter candidate.
\end{abstract}
\section{Introduction}
\begin{figure}
\centering
\begin{tikzpicture}[very thick,rounded corners=0.5pt,line cap=round,scale=0.53]
\shadedraw[top color=gray!20,bottom color=gray!70,yslant=0.1]
(0,0) -- (2,2) -- node[above=-13pt,rotate=-90]{UV-brane} (2,7) -- (0,5) --  cycle;
\shadedraw[top color=gray!25,bottom color=gray!75,yslant=0.1](8,0.5) -- (9,1.5) -- node[above,rotate=-90]{IR-brane} (9,4) -- (8,3) -- cycle;
\draw[thick,black,yslant=0.1,opacity=0.5](0,5) to [out=-30,in=170]node[above right,opacity=1]{$e^{-2k|y|}$} (8,3)
                                (2,7) to [out=-50,in=170] (9,4)
                                (2,2) to [out=8,in=180] (9,1.5)
                                (0,0) to [out=20,in=175] (8,0.5);
\draw (0,0) node[below]{$y=0$}
 (8,0.5)node[above]{$y=L$}
 (5,2.5)node[above,black]{$\Lambda_B$};
\draw[->,>=stealth,thick,yslant=0.1] (1,1) -- (1,2.1) node[above]{$x^\mu$};
\draw[->,>=stealth,thick,yslant=0.1](1,1)-- (2.1,1) node[right]{$y$};
\begin{scope}[xshift=15cm,yshift=3cm]
\draw[black] (0,0)circle (2cm);
\draw[<-,>=stealth,thick,gray] (0,2) -- (0,1.7);
\draw[<-,>=stealth,thick,gray] (0,-2)node[below, black]{$S_1/\mathbb{Z}_2$ orbifold} -- (0,-1.7);
\draw[<->,>=stealth,thick,gray] (1,1.7) -- (1,-1.7);
\draw[<->,>=stealth,thick,gray] (-1,1.7) -- (-1,-1.7);
\draw[thick, gray] (0,1)node[above]{$+y$} -- (0,-1)node[below]{$-y$};
\filldraw [gray] (-2,0)node[left,black]{$0$} circle (2pt);
\filldraw [gray] (2,0)node[right,black]{$L$} circle (2pt);
\end{scope}
\end{tikzpicture}
\caption{Cartoon of RS1 geometry.}
\label{fig_rs_geometry}
\end{figure}
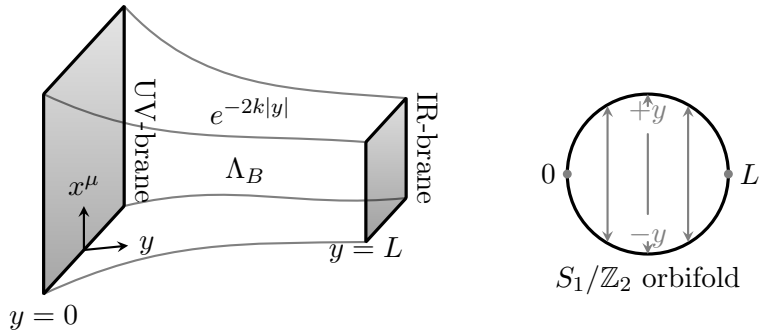

The 5D warped model of Randall and Sundrum (RS) with two D3-branes (RS1) provides an elegant solution to {\it the hierarchy problem} \cite{Randall:1999ee}. The two D3-branes are localized on the fixed points of the orbifold $S_1/\mathbb{Z}_2$, a ``UV-brane'' at $y=0$ and  an ``IR-brane'' at $y=L$ --- the UV-IR model, see Fig.~\ref{fig_rs_geometry}.
The solution for the RS geometry is \cite{Randall:1999ee},
\beq
ds^2=e^{2k|y|}\eta_{\mu\nu}dx^\mu dx^\nu+dy^2,  \label{metric}
\eeq
where $k$ is a constant of the order of 5D Planck mass $M_\ast$. Randall and Sundrum showed that if the 5D theory involves only one mass scale
$M_\ast$ then, due to the presence of non-trivial warping along the extra-dimension, the effective mass scale on the IR-brane is rescaled to $m_{KK}\equiv ke^{-kL}\sim\co(\text{TeV})$ for $kL\sim\co(37)$. However, the RS proposal \cite{Randall:1999ee} lacked the mechanism of stabilizing the separation $L$ between the two branes. A stabilization mechanism for the RS1 geometry was proposed by Goldberger and Wise (GW) \cite{Goldberger:1999uk}; which employs a real scalar field in the bulk of RS geometry with localized potentials on both branes.

The aim of this work is twofold: ({\it i}) to analyse the GW stabilization mechanism with an $SU(2)$ bulk Higgs doublet~\footnote{Recently Geller et al.~\cite{Geller:2013cfa} also treated an $SU(2)$ bulk Higgs doublet as a stabilization field, however, they assumed weak backreaction and have not solved the full Higgs-gravity coupled Einstein equations.} and, ({\it ii}) to investigate the lowest odd KK-mode of the bulk Higgs in a 5D $\mathbb{Z}_2$-symmetric model with {\it warped KK-parity} \cite{Ahmed:2015ona,Ahmed:Planck}. In order to achieve the first goal, we consider an $SU(2)$ Higgs doublet in the bulk of RS1 geometry in Sec.~\ref{Radius stabilization with a bulk Higgs doublet} and use the {\it superpotential method} to solve the Higgs-gravity coupled Einstein equations \cite{Ahmed:2015ona,Ahmed:2015}. The second goal is achieved  in Sec.~\ref{Warped Higgs dark matter}, where we introduce the IR-UV-IR model with a warped KK-parity and consider the 5D SM bosonic sector in the bulk of this model. In the zero-mode effective theory we have a stable scalar, which is the dark matter candidate, along with all the SM fields.

\section{Radius stabilization with a bulk Higgs doublet}
\label{Radius stabilization with a bulk Higgs doublet}
We employ an $SU(2)$ Higgs doublet $H$ in the bulk of a 5D geometric interval $y\in[0,L]$ with the following scalar-gravity action,
\begin{align}
S=\int d^5x \sqrt{-g}\Big\{2\mst^3R-\left\vert D_M H\right\vert^2 -V(H)-\sum_{i} V_i(H)\delta(y-y_i)\Big\},    \label{action}
\end{align}
where $V(H)$ and $V_{i}(H)$ (for  $i=1,2$) are the bulk and brane potentials, respectively, whereas $y_{1(2)}\equiv0(L)$ are the UV(IR)-brane locations. We employ the following metric ansatz:
\beq
ds^2=e^{2\sigma(y)}\eta_{\mu\nu}dx^\mu dx^\nu+dy^2, \label{metric}
\eeq
where $\sigma(y)$ is a $y$-dependent warp-function and $\eta_{\mu\nu}\equiv\text{diag}(-1,1,1,1$). We write the $y$-dependent vacuum expectation value (vev) of the Higgs field as:
\beq
\langle H\rangle=\frac{1}{\sqrt2}\bpm 0\\  h_{v}(y) \epm.	\label{higgs_vev}
\eeq
The background scalar-gravity coupled Einstein equations following from the action \eqref{action} and the metric ansatz \eqref{metric} can be solved analytically by using the {\it superpotential method} \cite{DeWolfe:1999cp}. We assume the scalar potential $V( h_{v})$ can be written in the following form:
\begin{equation}
V( h_{v})=\frac{1}{8}\left( \frac{\partial {\cal W}( h_{v})}{\partial h_{v}}\right)^{2}-\frac{1}{24\mst^3}{\cal W}( h_{v})^{2},
\label{potential}
\end{equation}
where the superpotential ${\cal W}( h_{v})$ satisfies:
\begin{align}
  h_{v}^{\prime}&=\frac12\frac{\partial {\cal W}( h_{v})}{\partial  h_{v}},  &\sigma^{\prime}&=-\frac{{\cal W}( h_{v})}{24\mst^3},
 \label{super_potential_eqs}\\
\frac12{\cal W}( h_{v})\Big\vert_{y_i-\epsilon}^{y_i+\epsilon}&=V_i( h_{v})\Big\vert_{ h_{v}= h_{v}(y_i)}, &\frac12\frac{\partial {\cal W}( h_{v})}{\partial h_{v}}\Big\vert_{y_i-\epsilon}^{y_i+\epsilon}&=\frac{\partial V_i( h_{v})}{\partial h_{v}}\Big\vert_{ h_{v}= h_{v}(y_i)} ,  \label{jump_conditions_W}
\end{align}
where a {\it prime} denotes a derivative w.r.t. the $y$-coordinate. The last line above follows from the {\it jump} conditions across the branes.

\begin{figure}
\begin{center}
\includegraphics[width=0.47\textwidth]{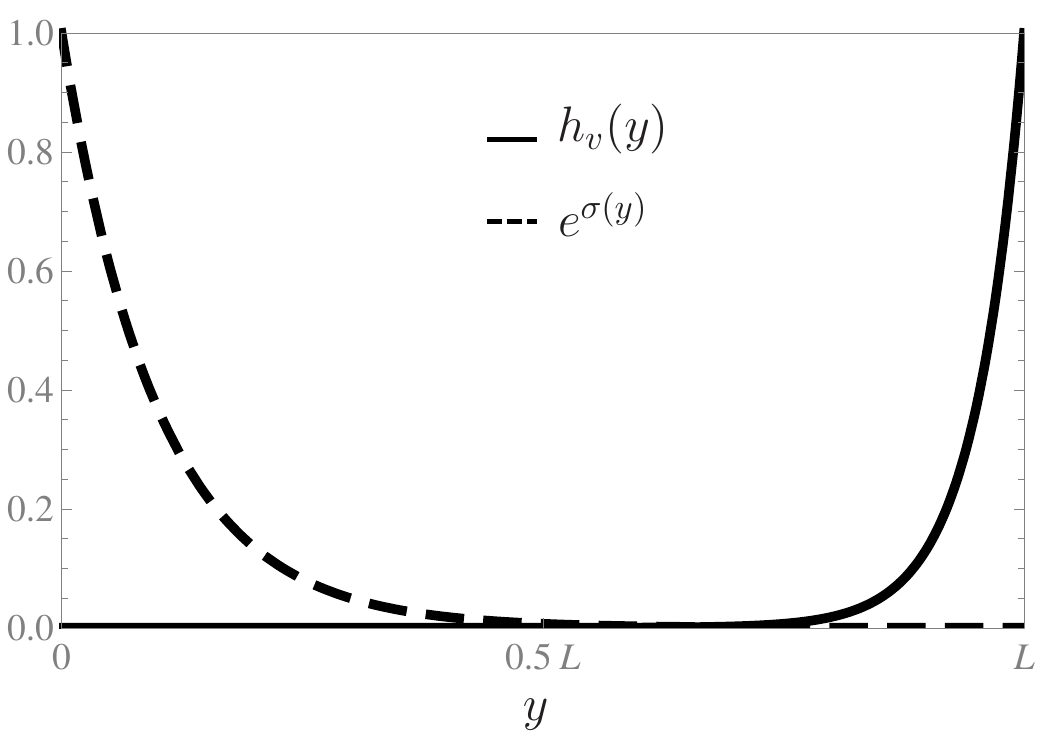}
\includegraphics[width=0.47\textwidth]{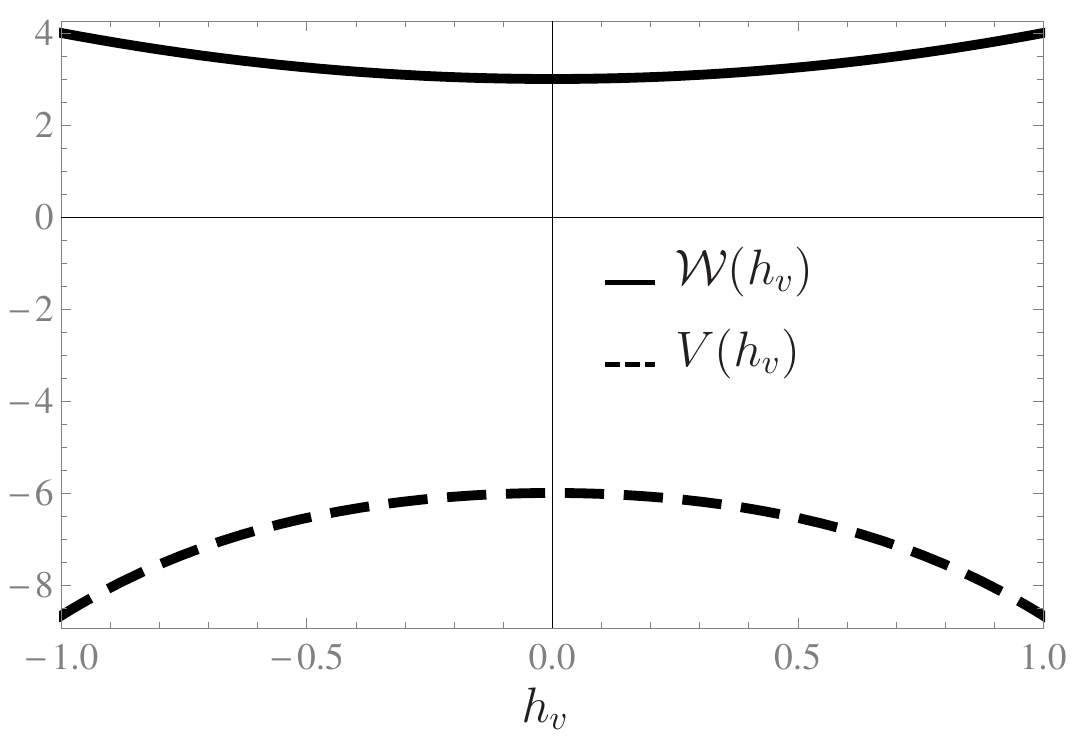}
\end{center}
\vspace{-7pt}
\caption{The left graph illustrates shapes of the background vev $h_{v}(y)$ and the warp factor $e^{\sigma(y)}$ as a function of $y$, whereas, the right graph shows shapes of superpotential ${\cal W}( h_{v})$ and the bulk potential $V( h_{v})$ as a function of background vev $ h_{v}$. The parameter choice adopted for the graphs is: $\Delta=2$ and $k=v_2=\mst=1$.}
\label{phiAWV}
\end{figure}

We consider the following form of superpotential ${\cal W}( h_{v})$
\beq
{\cal W}( h_{v})=24\mst^3k+\Delta k h_{v}^2, \Lsp  \Delta\equiv2+\sqrt{4+\mu_B^2/k^2}, \label{superpotential}
\eeq
where $\Delta$ parameterises the bulk Higgs mass $\mu_B$ and in the dual CFT it corresponds to the scaling dimension of the composite operator $\co_H$. The scalar potential $V( h_{v})$ following from \eq\eqref{potential} takes the form
\beq
V( h_{v})=-24\mst^3k^2+\frac12\mu_B^2 h_{v}^2-\frac{\Delta ^2k^2}{24\mst^3} h_{v}^4. \label{bulk_potenial}
\eeq
We employ the following forms of the brane-localized potentials,
\begin{align}
V_{1(2)}( h_{v})&=\pm{\cal W}(h_v)+\frac{\lambda_{1(2)}}{4k^2}\big( h_{v}^2- v_{1(2)}^2\big)^2,	\label{Vuv_ir}
\end{align}
where $ v_{1(2)}$ and $\lambda_{1(2)}$ are the values of background vevs  and quartic couplings at the UV (IR) branes, respectively. The background vev $ h_{v}(y)$ and warp-function $\sigma(y)$ are obtained by integrating Eq.~\eqref{super_potential_eqs} (see Fig.~\ref{phiAWV}):
\begin{align}
 h_{v}(y)= v_{2}e^{\Delta k(|y|-L)},    \lsp \sigma(y)=-k|y| -\frac{v_{2}^2e^{-2\Delta kL}}{48\mst^3} \Big[e^{2\Delta k|y|}-1\Big].  \label{warp_function_A}
\end{align}
The background vev profile $h_v(y)$ satisfies following normalization condition:
\beq
\int_{0}^{L}dy e^{2\sigma(y)} h_{v}^2(y)=v_{SM}^2,       \label{phi_vev_norm}
\eeq
where $v_{SM}\simeq246\gev$ is the SM Higgs vev.
The brane separation $L$ is fixed by \eq\eqref{warp_function_A} as:
\beq
kL=\frac{1}{\Delta}\ln\left(\frac{ v_{2}}{v_{1}}\right).
\eeq
For $\Delta\geq2$, in order to solve the gauge hierarchy problem one needs to have $kL\simeq\co(37)$, which implies that $v_1\ll v_2$~\footnote{Note that the requirement $v_1\ll v_2$ is not an extra fine-tuning other than the one required to set 4D cosmological constant zero, see also \cite{Geller:2013cfa}.}. For instance, \eq\eqref{phi_vev_norm} gives,
\beq
v_{SM}\simeq\frac{v_{2}e^{-kL}}{\sqrt{k(\Delta-1)}}\sim\co(\text{TeV}),	\label{phi_ir}
\eeq
for $v_{2}\sim\co(\mpl)$, $\Delta\geq2$ and $kL\sim\co(37)$.

\section{Warped Higgs dark matter}
\label{Warped Higgs dark matter}
In this section, we consider a $\mathbb{Z}_2$-symmetric, IR-UV-IR, warped extra dimensional model presented in Ref.~\cite{Ahmed:2015ona}. The geometric $\mathbb{Z}_2$-symmetry leads to a {\it warped KK-parity}, under which all the bulk fields have towers of even and odd KK-modes. The lowest odd KK-particle (LKP) is stable and hence can serve as a dark matter candidate.

As it is shown in Ref.~\cite{Ahmed:Planck} the ``IR-UV-IR'' geometry is equivalent to a single copy of ``UV-IR'' RS1 geometry, where all the bulk fields satisfy Neumann (or mixed) and Dirichlet boundary conditions (b.c.) at $y=0$ corresponding to the even and odd fields. Hence, we discuss merely UV-IR geometry with two sets of b.c. at $y=0$. We consider the bosonic sector of the SM gauge group in the bulk of RS1 geometry~\footnote{In this section we consider $\sigma(y)\simeq -k|y|$, i.e. weak back reaction. Moreover, we neglect the quartic terms in the bulk and UV-brane Higgs potentials, as these terms are suppressed, and introduce some convenient notations for brane potential parameters.},
\begin{align}
S=-\int d^5x \sqrt{-g}\bigg\{&\frac14 F^{a}_{MN}F^{aMN}+\frac14 B_{MN}B^{MN}+\left\vert D_M H\right\vert^2 +\mu_B^2|H|^2\notag\\
 &+V_{1}(H)\delta(y)+V_{2}(H)\delta(y-L)\bigg\},    \label{action_SM}
\end{align}
where the brane-localized potentials are
\beq
V_{1}(H)=\frac{m^2_{UV}}{k}|H|^2,   \Lsp V_{2}(H)=-\frac{m^2_{IR}}{k}|H|^2+\frac{\lambda_{IR}}{k^2}|H|^4,
\label{boudary_potentials}
\eeq
and $F^a_{MN}[B_{MN}]$ is the 5D field strength tensor for $SU(2)[U(1)_Y]$.

It is a straightforward exercise to obtain an effective 4D theory by integrating the 5th dimension after KK-decomposition of the bulk fields. After obtaining the 4D effective theory, one can get a low-energy 4D effective theory by assuming the KK-scale, $m_{KK}\equiv ke^{-kL}\sim\co(\text{few})\tev$, is much heavier than all the zero-modes of the theory  \cite{Ahmed:2015ona,Ahmed:Planck}. It turns out, after applying the appropriate b.c., that the odd zero-mode wave functions for the gauge fields (same for the odd Goldstone zero-modes) are zero. Hence the odd zero-mode gauge fields and the odd Goldstone modes of the zero-mode odd Higgs doublet are not present in the effective theory. With these observations, we can write the 4D effective Lagrangian for the zero-modes as,
\begin{align}
{\cal L}^{4D}_{eff}=&-\frac12 {\cal W}^{+}_{\mu\nu}{\cal W}^{-\mu\nu}- \frac14 {\cal Z}_{\mu\nu}{\cal Z}^{\mu\nu}-\frac14 {\cal F}_{\mu\nu}{\cal F}^{\mu\nu}- m^2_W W^+_\mu W^{-\mu}-\frac12 m^2_Z Z_\mu Z^\mu\notag\\
&-\frac12\partial_\mu h\partial^\mu h- \frac12m^2_h h^2  -\frac12\partial_\mu \chi\partial^\mu \chi- \frac12m^2_\chi \chi^2-\lambda vh^3-\frac\lambda4 h^4 \notag\\
&-\frac\lambda4 \chi^4-3\lambda v h\chi^2-\frac32\lambda h^2\chi^2-\frac{g_4^2}{2}vW_\mu^+W^{-\mu}h -\frac{g_4^2}{4}W_\mu^+W^{-\mu}(h^2+\chi^2)  \notag\\
&-\frac14(g_4^2+g_4^{\p2})vhZ_\mu Z^\mu  -\frac18(g_4^2+g_4^{\p2})Z_\mu Z^\mu (h^2+\chi^2),    \label{eff_lag}
\end{align}
where ${\cal V}_{\mu\nu}$ is the field strength tensor of the SM gauge bosons $V_\mu$. Notice that the above zero-mode 4D effective Lagrangian contains all the SM fields (including the Higgs boson $h$) plus a scalar $\chi$, component of the odd zero-mode Higgs field. The above Lagrangian has a $\mathbb{Z}_2$-symmetry~\footnote{This $\mathbb{Z}_2$ symmetry is a manifestation of the warped KK-parity, see e.g. \cite{Ahmed:2015ona,Ahmed:Planck}.}, under which all the SM fields are even, whereas the scalar $\chi$ is odd, i.e. $\chi\to-\chi$. Hence the scalar $\chi$ is our dark matter candidate.
The masses of the SM Higgs boson $h$, dark-Higgs $\chi$ and gauge bosons are
\begin{align}
m^2_h&=2\mu^2, \lsp m^2_\chi=m_h^2+\delta m^2, \lsp m^2_{W}=\frac{g_4^2m^2_{Z}}{g^2_4+g^{\p2}_4}=\frac{1}{4}g^2_4\frac{\mu^2}{\lambda},         \label{masses_higgs_WZ}
\end{align}
where $m_h=125\gev$ and $\delta m^2\equiv 3m_{KK}^2m_t^2/{(4\pi^2v_{SM}^2)}$ is the shift in the dark-Higgs mass due to quantum correction (quadratically divergent) below the cut-off scale $m_{KK}$ \cite{Ahmed:2015ona}. The parameters of the above effective Lagrangian $\mu,~v_{SM},~\lambda,~g_4$ and $g^\p_4$ are determined in terms of the 5D fundamental parameters of the theory, which are then fixed in such a way that we recover all the SM values of these parameters within our low-energy effective theory.

\begin{figure}
\centering
\begin{tikzpicture}[node distance=1cm,very thick, rounded corners=0pt,line cap=round,scale=0.9,transform shape]
\begin{scope}[xshift=0.5cm]
\coordinate[] (v1);
\coordinate[above left=of v1] (a1);
\coordinate[below left=of v1] (a2);
\coordinate[above right=of v1] (b1);
\coordinate[below right=of v1] (b2);
\draw[scalar] (a1)node[left]{$\chi$} -- (v1);
\draw[scalar] (a2)node[left]{$\chi$} -- (v1);
\draw[boson](v1)--(b1)node[right]{$W,Z$};
\draw[boson](v1)--(b2)node[right]{$W,Z$};
\filldraw [gray] (v1) circle (2pt);
\end{scope}
\begin{scope}[xshift=4.5cm]
\coordinate[] (v1);
\coordinate[above left=of v1] (a1);
\coordinate[below left=of v1] (a2);
\coordinate[right=1.3cm of v1] (v2);
\coordinate[above right=of v2] (b1);
\coordinate[below right=of v2] (b2);
\draw[scalar] (a1)node[left]{$\chi$} -- (v1);
\draw[scalar] (a2)node[left]{$\chi$} -- (v1);
\draw[scalar] (v1)--node[above]{$h$}(v2);
\draw[boson](v2)--(b1)node[right]{$W,Z$};
\draw[boson](v2)--(b2)node[right]{$W,Z$};
\filldraw [gray] (v1) circle (2pt)
                (v2) circle (2pt);
\end{scope}
\begin{scope}[xshift=9.5cm]
\coordinate[] (v1);
\coordinate[above left=of v1] (a1);
\coordinate[below left=of v1] (a2);
\coordinate[right=1.3cm of v1] (v2);
\coordinate[above right=of v2] (b1);
\coordinate[below right=of v2] (b2);
\draw[scalar] (a1)node[left]{$\chi$} -- (v1);
\draw[scalar] (a2)node[left]{$\chi$} -- (v1);
\draw[scalar] (v1)--node[above]{$h$}(v2);
\draw[fermion](v2)--(b1)node[right]{$f$};
\draw[fermion](b2)node[right]{$\bar f$} -- (v2);
\filldraw [gray] (v1) circle (2pt)
                (v2) circle (2pt);
\end{scope}\newline
\begin{scope}[yshift=-2.2cm]
\coordinate[] (v1);
\coordinate[above left=of v1] (a1);
\coordinate[below left=of v1] (a2);
\coordinate[above right=of v1] (b1);
\coordinate[below right=of v1] (b2);
\draw[scalar] (a1)node[left]{$\chi$} -- (v1);
\draw[scalar] (a2)node[left]{$\chi$} -- (v1);
\draw[scalar](v1)--(b1)node[right]{$h$};
\draw[scalar](v1)--(b2)node[right]{$h$};
\filldraw [gray] (v1) circle (2pt);
\end{scope}
\begin{scope}[xshift=3cm,yshift=-2.2cm]
\coordinate[] (v1);
\coordinate[above left=of v1] (a1);
\coordinate[below left=of v1] (a2);
\coordinate[right=1.3cm of v1] (v2);
\coordinate[above right=of v2] (b1);
\coordinate[below right=of v2] (b2);
\draw[scalar] (a1)node[left]{$\chi$} -- (v1);
\draw[scalar] (a2)node[left]{$\chi$} -- (v1);
\draw[scalar] (v1)--node[above]{$h$}(v2);
\draw[scalar](v2)--(b1)node[right]{$h$};
\draw[scalar](v2) -- (b2)node[right]{$h$};
\filldraw [gray] (v1) circle (2pt)
                (v2) circle (2pt);
\end{scope}
\begin{scope}[xshift=7.5cm,yshift=-1.7cm]
\coordinate[] (v1);
\coordinate[below= of v1] (v2);
\coordinate[position=150 degrees from v1] (a1);
\coordinate[position=-150 degrees from v2] (a2);
\coordinate[position=30 degrees from v1] (b1);
\coordinate[position=-30 degrees from v2] (b2);
\draw[scalar] (a1)node[left]{$\chi$} -- (v1);
\draw[scalar] (a2)node[left]{$\chi$} -- (v2);
\draw[scalar] (v1)--node[left]{$\chi$}(v2);
\draw[scalar](v1)--(b1)node[right]{$h$};
\draw[scalar](v2)--(b2)node[right]{$h$};
\filldraw [gray] (v1) circle (2pt)
                (v2) circle (2pt);
\end{scope}
\begin{scope}[xshift=11cm,yshift=-1.7cm]
\coordinate[] (v1);
\coordinate[below= of v1] (v2);
\coordinate[position=150 degrees from v1] (a1);
\coordinate[position=-150 degrees from v2] (a2);
\coordinate[above right=1.8cm of v2] (b1);
\coordinate[below right=1.8cm of v1] (b2);
\draw[scalar] (a1)node[left]{$\chi$} -- (v1);
\draw[scalar] (a2)node[left]{$\chi$} -- (v2);
\draw[scalar] (v1)--node[left]{$\chi$}(v2);
\draw[scalar](v2)--(b1)node[right]{$h$};
\draw[scalar](v1)--(b2)node[right]{$h$};
\filldraw [gray] (v1) circle (2pt)
                (v2) circle (2pt);
\end{scope}
\end{tikzpicture}
\caption{Dark matter annihilation diagrams.}
\label{DM_annihi_diagrams}
\end{figure}
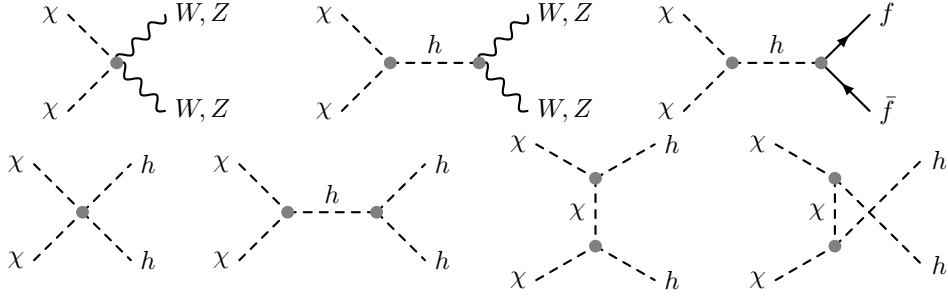

\begin{figure}
\begin{center}
\includegraphics[width=0.5\textwidth]{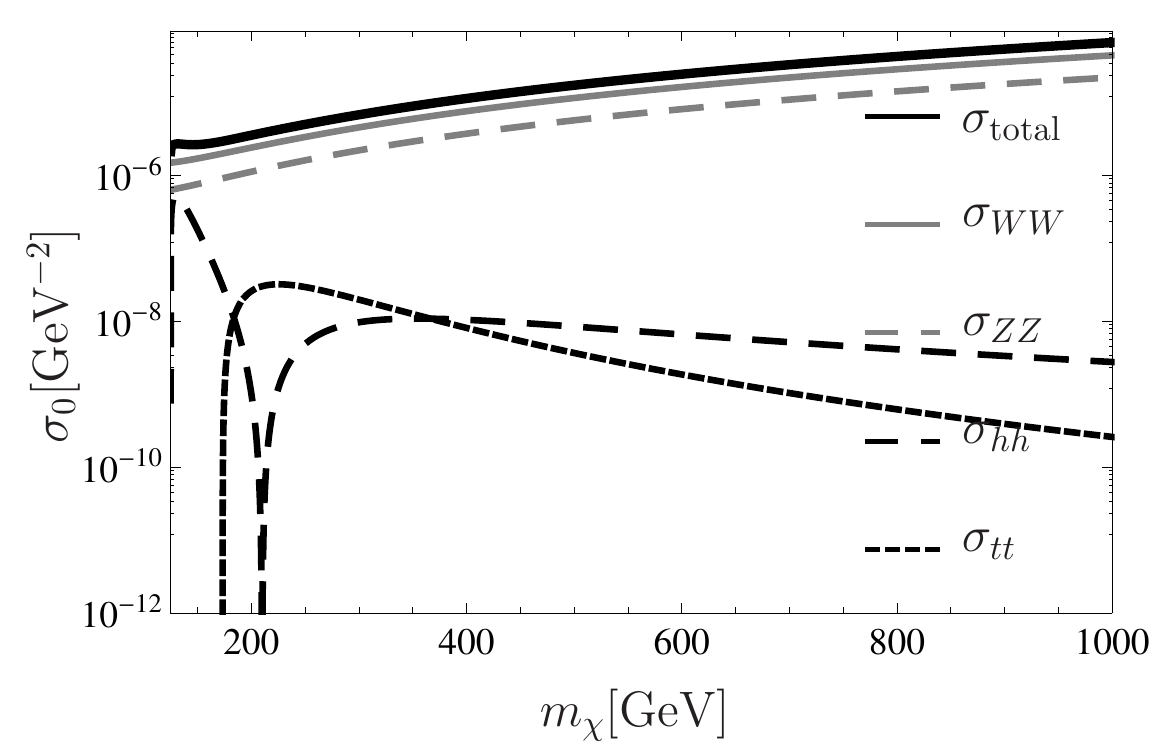}
\includegraphics[width=0.47\textwidth]{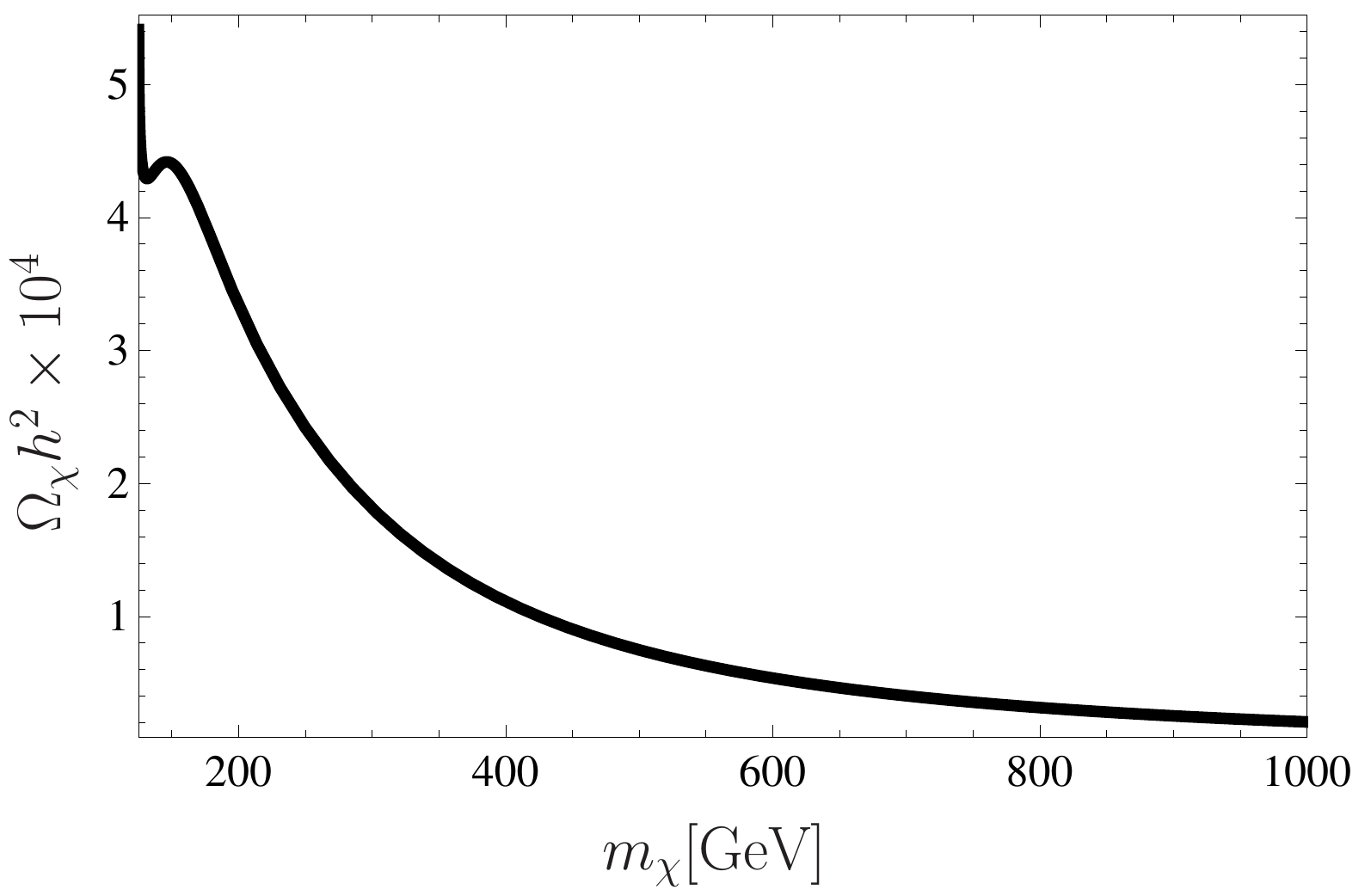}
\end{center}
\vspace{-7pt}
\caption{The above graphs show the annihilation cross-section $\sigma_0$ for different final states (left)
and the $\chi$ abundance $\Omega_\chi h^2\times 10^{4}$ (right) as a function of dark matter mass $m_\chi$.}
\label{sigma_omega}
\end{figure}

Now we calculate the annihilation cross-section and the relic abundance of the dark-matter. The Feynman diagrams contributing to dark matter annihilation are shown in Fig.~\ref{DM_annihi_diagrams}.
In Fig.~\ref{sigma_omega} (left panel) we have plotted the annihilation cross-section for
the contributing channels as a function of $m_\chi$. As shown in the graph the total cross section is dominated by $WW$ and $ZZ$ final states. The main contributions for these final states are those generated by contact interactions $\chi\chi WW(ZZ)$, whereas, all the other final states that include the Higgs boson $h$ or the top quark is very small in comparison to $\chi\chi \to WW(ZZ)$. The dark matter relic abundance $\Omega_\chi h^2$ is shown in Fig.~\ref{sigma_omega} (right panel). We observe that $\Omega_\chi h^2\lsim 10^{-4} $ once the electroweak precision bound on the KK mass scale $m_{KK}$ is imposed \cite{Ahmed:2015ona}.


\section{Conclusions}
\label{Conclusions}
In this work we have investigated two implications of the bulk Higgs doublet in warped extra dimensions.
\bit
\item We present the $SU(2)$ bulk Higgs doublet in warped extra dimension as the GW stabilizing field, where the superpotential method is employed to solve the Higgs-gravity coupled Einstein equations. We show that an IR-localized Higgs background can fix the size of extra dimension such that it solves the hierarchy problem.
\item We consider a geometric $\mathbb{Z}_2$ symmetric model of warped extra dimension, the IR-UV-IR, which has a warped KK-parity. Within this model we investigated the zero-mode effective theory for the bulk Standard Model (SM) bosonic sector. This effective theory contains all the SM fields and a stable scalar particle which is a dark matter candidate. It is found that this dark matter candidate can provide only a small fraction of the observed dark matter abundance.
\eit

This work has been supported in part by the National Science Centre
(Poland) as  research projects no DEC-2014/15/B/ST2/00108
and DEC-2014/13/B/ST2/03969. AA would like to thank Barry Dillon for reading the draft and valuable comments.


\end{document}